\documentstyle[12pt,aps]{revtex}
\begin{document}

\def\PHI{\hbox{$\underline{\phi}$}}
\def\PI{\hbox{$\underline{\pi}$}}
\def\RHO{\hbox{$\underline{\rho}$}}
\def\TAU{\hbox{$\underline{\tau}$}}
\def\As{\hbox{\rm\rlap/A}}
\def\Cs{\hbox{\rm\rlap/C}}
\def\ds{\hbox{\rm\rlap/$\partial$}}
\def\Ds{\hbox{\rm\rlap/D}}
\def\Gs{\hbox{\rm\rlap/G}}
\def\omegas{\hbox{\rm\rlap/$\omega$}}
\def\thetas{\hbox{\rm\rlap/$\theta$}}
\def\vphis{\hbox{\rm\rlap/$\varphi$}}
\def\up{\hbox{$\partial^{\mu}$}}
\def\down{\hbox{$\partial_{\mu}$}}
\def\L{\hbox{$\cal L$}}
\def\O{\hbox{$\cal O$}}

\title{
\begin{flushright}
\vspace{-1cm}
{\normalsize MC/TH 95/14}
\vspace{1cm}
\end{flushright}
$\rho \to 4\pi$ in chirally symmetric models\\
\vskip 10pt}
\author{Robert S. Plant and Michael C. Birse}
\address{Theoretical Physics Group, Department of Physics and Astronomy,\\
University of Manchester, Manchester, M13 9PL, U.K.\\}
\maketitle
\vskip 20pt

\begin{abstract}
The decays $\rho^0\to 2\pi^+2\pi^-$ and $\rho^0\to 2\pi^0\pi^+\pi^-$ are
studied using various effective Lagrangians for $\pi$ and $\rho$ (and in some
cases $a_1$) mesons, all of which respect the approximate chiral symmetry of
the strong interaction. Partial widths of the order of 1 keV or less are found
in all cases. These are an order of magnitude smaller than recent predictions
based on non-chiral models.
\end{abstract}
\vskip 20pt

There have been recent hopes\cite{BGP93,ESK95} that the rare decays
$\rho^{0} \to 2\pi^+2\pi^-$ and $\rho^{0} \to 2\pi^0\pi^+\pi^-$ might be
observable in the near future in experiments at high-luminosity $e^{+}e^{-}$
machines, such as VEPP-2M\cite{Do+91} or DA$\Phi$NE\cite{Pan94}. These decays
might test various aspects of the low-energy effective Lagrangians that have
been proposed for the interactions of $\rho$-mesons and
pions\cite{Mei88,Ec+89,Bir95}. For example, the $2\pi^0\pi^+\pi^-$ mode is
sensitive to the $3\rho$ coupling that appears in theories with gauge-type
couplings of the $\rho$. Current experimental limits on the partial widths for
the decays are 30 keV for the $2\pi^+2\pi^-$ mode\cite{Ku+88} and 6 keV for
the $2\pi^0\pi^+\pi^-$ mode\cite{Au+87}. These are already sufficient to rule
out some earlier estimates which did not incorporate chiral symmetry, for
example one dominated by $\pi a_1$ and $\pi a_2$ intermediate
states\cite{Par82}.

Bramon, Grau and Pancheri\cite{BGP93} studied the $2\pi^+2\pi^-$ decay with
two commonly-used formalisms for including the $\rho$ meson in low-energy
effective Lagrangians. For the simplest ``hidden-gauge" theory of Bando {\it
et al.}\cite{Ba+85,BKY88} they obtained a partial width of $7.5\pm 0.8$ keV.
In contrast, for a simplified ``massive Yang-Mills" theory, where the $\rho$ is
coupled to a sigma model as an SU(2)$_V$ gauge boson\cite{KRS84,Mei88}, they
obtained $60\pm 7$ keV. This suggested that the process could distinguish
between these approaches, and indeed that the massive Yang-Mills one is
already ruled out by the data.

More recently, Eidelman, Silagadze and Kuraev\cite{ESK95} have pointed out
that the massive Yang-Mills model of Ref.~\cite{BGP93} does not respect chiral
symmetry. One possibility is to use gauge couplings corresponding to the full
SU(2)$_R\times$SU(2)$_L$ symmetry by introducing the $a_1$ as the chiral
partner of the $\rho$\cite{KRS84,Mei88}. An alternative, and the one followed
in Ref.\cite{ESK95}, is that proposed by Brihaye, Pak and Rossi\cite{BPR85}
where additional terms are added to the naive $\pi$-$\rho$ Lagrangian in order
to preserve low-energy theorems. Including $4\pi$ and $\rho 4\pi$ correction
terms, the authors of Ref.\cite{ESK95} obtained partial widths of $16\pm 1$
keV for the $2\pi^{+}2\pi^{-}$ mode and $6.0\pm 0.2$ keV for the
$2\pi^{0}\pi^{+}\pi^{-}$ mode, consistent with the present experimental limits
and within reach of future experiments.

Although the Lagrangian of Ref.\cite{ESK95} is constructed to satisfy
vector-meson dominance and a range of low-energy theorems for $\pi\pi$
scattering and the couplings of photons, it is still not chirally symmetric. A
Yang-Mills Lagrangian for pions and $\rho$ mesons only contains a
momentum-independent $2\pi2\rho$ coupling of the form
$${\frac {1}{2}} g_{\rho\pi\pi}^{2} (\PI\times\RHO_{\mu})^{2}. \eqno(1)$$
This leads to a $\pi$-$\rho$ scattering amplitude that does not vanish at
threshold in the chiral limit, and so is in conflict with low-energy theorems
of chiral symmetry\cite{DGH92}. One can add further terms to such a Lagrangian
to cancel out this unwanted behaviour, as described in Ref.~\cite{Bir95}.
It is, however, straightforward to work with a manifestly chiral approach from
the start. We have therefore calculated the $\rho^0\to 4\pi$ decay rates using
several such Lagrangians.

The low-energy interactions among pions can be described by a chirally
symmetric sigma model. There is, however, considerable freedom in extending
such a description to incorporate spin-1 mesons\cite{Mei88,Ec+89,Bir95}. We
have examined three such approaches: i) the hidden-gauge theories of Bando
{\it et al.}\cite{Ba+85,BKY88} in which a local SU(2) symmetry is introduced
into the non-linear sigma model and the $\rho$ is treated as the gauge boson of
this symmetry; ii) the massive Yang-Mills theories in which $\rho$ and $a_1$
fields transform linearly under the chiral symmetry; iii) the approach
suggested by Weinberg\cite{Wei68} and developed by Callan, Coleman, Wess and
Zumino\cite{CCWZ69} (CCWZ) in which the spin-1 mesons transform homogeneously
under a non-linear realisation of chiral symmetry. One should remember that,
in general, all of these are equivalent and any Lagrangian of one approach can
be converted into a corresponding Lagrangian of
another\cite{Mei88,Ec+89,Bir95}.

If we work at tree level, the relevant diagrams are shown in Figure~1.
Including anomalous terms in the effective Lagrangian leads to additional
contributions, similar to those of 1(f) and 1(g) but involving an intermediate
$\omega$ meson. In all cases we include a symmetry-breaking term of the form
$${\frac {f_{\pi}^{2}}{4}} m_{\pi}^{2} Tr(U + U^{\dagger}), \eqno(2)$$
where $U=\exp(i\TAU.\PI/f_{\pi})$.
This term produces the non-zero pion mass. It also has the effect of
modifying the $4\pi$ vertex in diagram 1(b). This is quantitatively important
for our results, since they are much smaller that those of
Refs.\cite{BGP93,ESK95}, but it does not affect their qualitative features.

First, we consider the simplest example of a hidden-gauge theory, defined by
the Lagrangian of Bando {\it et al.}\cite{Ba+85}. To allow for contributions
from the anomalous part of the Lagrangian, we introduce the $\omega$ meson by
enlarging the gauge group to $U(2)_{V}$. There are then three possible
gauge-covariant anomalous terms, with undetermined coefficients. We adopt the
suggestion of Ref.\cite{FKTUY85} with regard to these coefficients, including
an $\omega\rho\pi$ vertex but no $\omega3\pi$ contact term. The parameters of
this model satisfy the relations
$$m_{\rho}^{2} = ag^{2}f_{\pi}^{2},\qquad g_{\rho\pi\pi} ={\frac {a}{2}}g.
\eqno(3)$$
The choice $a=2$ yields the KSRF relation\cite{KSRF66}, universal coupling of
the $\rho$\cite{AFFR73} and allows vector dominance to be realised in the
$\gamma\pi\pi$ coupling. In our calculations we take the following values for
the parameters:
$$f_{\pi} = 92.4\hbox{MeV},\quad m_{\pi} = 139.6\hbox{MeV},\quad m_{\rho}
= m_{\omega} = 770\hbox{MeV},\quad a = 2,\quad g = 5.89.\eqno(4)$$

The amplitudes for the $\rho^0\to 4\pi$ decays are straightforward to derive,
but the final expressions are rather long and so we do not present them here.
Having found them, we must integrate over phase space to obtain the
corresponding partial widths. We express the integrals in terms of the
variables of Ref.\cite{Kum69} and evaluate them numerically using the NAG
routine D01FDF, which maps the region of integration onto an $n$-dimensional
sphere and uses the method of Sag and Szekeres\cite{SS64} to perform the
integration. With a suitable choice of the parameter controlling the mapping
onto the sphere, we find that about 50,000 integration points are sufficient to
give the integrals to an accuracy of one part in a thousand.

The results obtained in this model are shown in the first line of Table~1,
labelled HG. They are roughly an order of magnitude smaller than any of the
results given in Refs.\cite{BGP93,ESK95}. Bramon {\it et al.}\cite{BGP93} have
calculated the process $\rho^{0} \to 2\pi^{+}2\pi^{-}$ in the same model and
they quote a value of $7.5 \pm 0.8\hbox{keV}$, in sharp contrast with our
result. The crucial difference between the calculations lies in the strength
of the direct $\rho4\pi$ coupling. Bramon {\it et al.}\ assumed that the
expression for this vertex was the same as in a massive Yang-Mills model,
specifically the $\rho4\pi$ term of
$$-ig_{\rho\pi\pi}{f_\pi^2\over 2}\hbox{Tr}[\rho^\mu(U^\dagger\partial_\mu U
+U\partial_\mu U^\dagger)]=g_{\rho\pi\pi} \left(1 - {\frac {1}{3f_{\pi}^{2}}}
\PI^{2} + \cdots \right) \RHO_{\mu}.\PI\times\up\PI.\eqno(5)$$
In fact, the corresponding term in the hidden-gauge model should be written, in
the unitary gauge, as
$$-2iagf_\pi^2\hbox{Tr}[\rho^\mu(u^\dagger\partial_\mu u+u\partial_\mu
u^\dagger)]= {\frac {a}{2}} g \left(1 - {\frac {1}{12f_{\pi}^{2}}} \PI^{2}
+ \cdots \right) \RHO_{\mu}.\PI\times\up\PI. \eqno(6)$$
where $u$ is the square root of $U$. Although both expressions yield the same
$\rho\pi\pi$ coupling, the $\rho4\pi$ terms differ by a factor of four. Hence
one cannot take the latter coupling to be the same as in a massive Yang-Mills
model. Reducing the contribution of diagram 1(a) by a factor of four has a
large effect on the total amplitude, explaining the difference between the
result of Ref.\cite{BGP93} and ours.\footnote{Comparison of the amplitudes
given by Refs.\cite{BGP93,ESK95} also indicates a discrepancy in the
evaluation of diagram 1(b). Our evaluation concurs with that of
Ref.\cite{ESK95}. However, the impact of this error is fairly small.}

We have examined the importance of the anomalous processes for our results.
Keeping only the non-anomalous contributions leads to the result for $\rho^{0}
\to 2\pi^0\pi^+\pi^-$ labelled HGNA, showing that the two types of term are
of similar importance. We have also looked at the effect of omitting the
symmetry-breaking $4\pi$ interaction of Eq.~(2) (but still keeping the physical
pion mass in the propagator etc.). The results, labelled HGCS, indicate that
this term is a smaller, but still significant, contribution. As a simple
estimate of contributions beyond tree level, we have examined the effect of
including the finite width of the $\rho$ in its propagator (as in
Ref.\cite{ESK95}). We find that this does not alter the results significantly.

We now turn to the massive Yang-Mills type of theory, so named because the
spin-1 mesons are introduced as if they were gauge bosons of chiral symmetry.
Local chiral symmetry is broken by the mass terms for those mesons. As
mentioned above, chiral symmetry requires that the $a_1$ be included as the
chiral partner of the $\rho$ in these models. We start with the simplest
example of such a Lagrangian, which is just a version of that in
Refs.\cite{LN68,GG69} but using a non-linear rather than a linear sigma
model. In this minimal model the $\rho$ and $a_1$ masses satisfy Weinberg's
relation, $m_{a_{1}} = \sqrt{2} m_{\rho}$ \cite{Wei67}. Note that, in this
approach, chiral symmetry requires strong cancellations among the
contributions to any amplitude. One cannot replace by hand, say, the $a_1$
mass by its physical value in such a calculation without violating chiral
symmetry. The constraints imposed by the symmetry mean also that one must not
omit the $a_1$ field from this approach without introducing compensating terms
that maintain chiral low-energy theorems\cite{BPR85,Bir95}.

A feature of this approach is the appearance of a $\pi$-$a_1$ mixing term in
the Lagrangian. To remove this and diagonalise the free-field part of the
Lagrangian, one can subtract a term proportional to $\partial_\mu\PI$ from the
axial field and renormalise the pion field. This constitutes a minimal
diagonalization procedure and leads to physical fields with complicated
chiral-transformation properties, but it is sufficient for our present
purposes. The procedure generates various additional three- and four-point
interactions arising from the gauge-invariant kinetic energies of the spin-1
fields. These include a $\rho\pi\pi$ interaction of order $\O(p^3)$ which
reduces the $\rho\pi\pi$ coupling strength at $m_\rho$ by a factor of ${\frac
{3}{4}}$ compared to its value at zero momentum. The minimal model is thus
unable to give a good description of the $\rho$ and $a_1$ masses and widths.
To remove this deficiency, the model must be supplemented with additional
terms\cite{KRS84,Mei88}, a point we return to below.

For the moment though, we consider the minimal model with the physical value
of $m_\rho$ and $g=g_{\rho\pi\pi} = 5.89$. We extend the model to include
the $\omega$ meson, taking the relevant anomalous $\omega\rho\pi$ and
$\omega3\pi$ vertices from Ref.~\cite{KRS84}. The results for $\rho^0\to 4\pi$
are shown in Table~1, labelled MMYM. They are similar in magnitude to those of
the hidden-gauge model. As already mentioned, the calculations of
Refs.\cite{BGP93,ESK95} used Yang-Mills models that do not respect chiral
symmetry. Without the ensuing cancellations, they lead to partial widths that
are an order of magnitude larger than ours.

Finally we consider several Lagrangians of the CCWZ type\cite{CCWZ69,Ec+89}.
These are expressed in terms of spin-1 fields that transform homogeneously
under a non-linear realisation of chiral symmetry\cite{Wei68,CCWZ69}. Our
reasons for this are twofold. First, the models just described can be
converted by a change of variables into equivalent CCWZ
Lagrangians\cite{Ec+89,Bir95}, which should yield the same predictions for any
observable as the original models. These therefore provide a useful check on
our results above. Second, the CCWZ formalism provides a convenient framework
to examine the sensitivity of our results to assumptions about the $a_1$. In
contrast to the massive Yang-Mills approach, the parameters describing the
$a_1$ mass and couplings may be independently changed without the need to
introduce further terms into the Lagrangian.

Apart from labelling of the coefficients, we use the notation of
Ref.\cite{Bir95} to express the relevant terms in the non-anomalous Lagrangian
as:
$$\L = {\frac {f_{\pi}^{2}}{4}} Tr (u_{\mu}u^{\mu}) + m_{\rho}^{2} Tr
(V_{\mu}V^{\mu}) + m_{a_{1}}^{2} Tr (A_{\mu}A^{\mu}) - {\frac {1}{2}} Tr
(V_{\mu\nu}V^{\mu\nu}) - {\frac {1}{2}} Tr (A_{\mu\nu}A^{\mu\nu})$$
$$\qquad - {\frac {i}{2}} g_{1} Tr (V_{\mu\nu}[u^{\mu},u^{\nu}])
+ {\frac {i}{2}} g_{2} Tr (V_{\mu\nu}[V^{\mu},V^{\nu}])
+ {\frac {i}{2}} g_{3} Tr \left( V_{\mu\nu} ([u^{\mu},A^{\nu}]
- [u^{\nu},A^{\mu}]) \right)$$
$$\qquad\qquad + {\frac {i}{2}} g_{4} Tr \left( A_{\mu\nu} ([u^{\mu},V^{\nu}]
- [u^{\nu},V^{\mu}]) \right) + {\frac {1}{8}} c_{1} Tr \left(
[u_{\mu},u_{\nu}]^{2} \right) - {\frac {1}{4}} c_{2} Tr \left(
[u_{\mu},u_{\nu}] [V^{\mu},V^{\nu}] \right)$$
$$ + {\frac {1}{8}} c_{3} Tr \left( ([u_{\mu},V_{\nu}] -
[u_{\mu},V_{\nu}])^{2} \right) - {\frac {1}{4}} c_{4} Tr \left(
[u_{\mu},u_{\nu}] ([u^{\mu},A^{\nu}]  - [u^{\nu},A^{\mu}]) \right).\eqno(7)$$
The minimal Yang-Mills model discussed above corresponds to the following set
of coefficients:
$$g_{1} = {\frac {1}{2g}} (1 - Z^{4}),\quad g_{2} = 2g,\quad g_{3} =
g_{4} = Z^{2},$$
$$c_{1} = g_{1}^{2},\quad c_{2} = 1 - Z^{4},\quad c_{3} = Z^{4},\quad
c_{4} = g_{1}g_{3}, \eqno(8)$$
where
$$Z^2=1-{g^2f_\pi^2\over m_\rho^2}. \eqno(9)$$
The appropriate coefficients for the simplest hidden-gauge model\cite{Ba+85}
can be obtained by setting $Z=0$ in Eq.~$(8)$ and dropping all terms involving
the $a_1$. We have calculated the amplitudes for $\rho^{0} \to 4\pi$ within
this framework and verified the results for the non-anomalous parts of the
hidden-gauge and massive Yang-Mills models described above.

The anomalous sectors of these models can also be converted into CCWZ form,
although this is somewhat involved for the massive Yang--Mills approach.
However the sum of amplitudes for the anomalous diagrams must remain
unaltered by the change of variables. For simplicity, therefore, we take the
anomalous part of the amplitude directly from the original version of the
massive Yang-Mills model. For the couplings given in Eqs.~(8,9), $m_\omega=783$
MeV and $m_{a_1}=1230$ MeV, we get the results labelled MYM$+1$ in Table~1.
These correspond to a massive Yang-Mills Lagrangian with non-minimal terms of
the type suggested in Refs.\cite{KRS84,Mei88}.

The parameter choice of Eqs.~(8,9) still suffers from the fact that it gives
too small a width for $\rho\to 2\pi$. It is straightforward to modify the CCWZ
parameters to remove this deficiency. This is equivalent to adding a further
non-minimal term to the massive Yang-Mills Lagrangian\cite{KRS84,Mei88} in
order to cancel the diagonalization-induced $\O(p^{3})$ $\rho\pi\pi$ coupling.
In the CCWZ representation, the corresponding parameters are
$$g_{1} = {\frac {1}{2g}},\qquad g_{2} = 2g,\qquad g_{3} = 0,\qquad
g_{4} = Z^{2},$$
$$c_{1} = {\frac {1}{4g^{2}}} (1 - Z^{8}),\qquad c_{2} = 1,\qquad
c_{3} = Z^{4},\qquad c_{4} = 0.\eqno(10)$$
The results for this set, with the empirical meson masses, are labelled MYM$+2$
in Table~1.

We have examined the sensitivity of our results to the $3\rho$ coupling
present in these Lagrangians. In the hidden-gauge and Yang-Mills models
described above this coupling is equal to the $\rho\pi\pi$ one because of the
assumed universal coupling of the $\rho$. Using the CCWZ equivalents of these
models, we have varied $g_2$ by $\pm 30\%$ and found that the decay rate for
$\rho^{0} \to 2\pi^0\pi^+\pi^-$ changes by about $\pm 1\%$.

These results show that, for all of the chirally symmetric models considered,
the partial widths for the decays $\rho^{0} \to 2\pi^+2\pi^-$ and $\rho^{0}
\to 2\pi^0\pi^+\pi^-$ are of the order of 1 keV, corresponding to cross
sections of the order of 5 pb. These are an order of magnitude smaller than
the predictions of Refs.\cite{BGP93,ESK95}. Although the processes may be hard
to observe in future experiments, they should not be beyond the reach of
DA$\Phi$NE which is designed to have a luminosity of $5\times 10^8$
b$^{-1}$s$^{-1}$\cite{Pan94}. The precise results are sensitive to the choice
of Lagrangian and parameters, receiving significant contributions from
anomalous processes (about which there is still some debate\cite{KS95}) and
symmetry-breaking interactions. However, any uncertainties in the
determination of the relevant parameters are not sufficient to affect our
general conclusion on the magnitudes of the decay rates for for $\rho^0\to
4\pi$ .

\section*{Acknowledgements}
We are grateful to Z. Silagadze for correspondence on these matters and to J.
McGovern for critically reading the manuscript. This work is supported by the
EPSRC and PPARC.

\newpage

\section*{Table}

\begin{center}
\begin{tabular}{| l | c | c |}
\hline
Model&\ $\rho^{0} \to 2\pi^{+} 2\pi^{-}$\ &\ $\rho^{0} \to
2\pi^{0} \pi^{+} \pi^{-}$ \\
\hline
HG&$0.89$&$0.44$\\
HGNA&$0.89$&$0.24$\\
HGCS&$0.59$&$0.37$\\
MMYM&$0.68$&$0.37$\\
MYM$+1$&$0.63$&$0.34$\\
MYM$+2$&$1.03$&$0.39$\\
\hline
\end{tabular}
\end{center}
\noindent Table 1. Decay widths (in keV) for $\rho^0\to 4\pi$ in various
versions of the chiral models. See the text for the definitions of these.

\newpage

\section*{Figure}

\setlength{\unitlength}{1mm}
\begin{picture}(140,30)
\put(5,15){\line(1,0){15}}
\put(5,16){\line(1,0){15}}
\put(20,15){\line(1,1){10}}
\put(20,15){\line(2,1){10}}
\put(20,16){\line(1,-1){10}}
\put(20,16){\line(2,-1){10}}
\put(15,3){(a)}

\put(70,15){\line(1,0){15}}
\put(70,16){\line(1,0){15}}
\put(85,15){\line(1,1){10}}
\put(85,16){\line(1,-1){8}}
\put(93,8){\line(1,0){10}}
\put(93,8){\line(2,1){10}}
\put(93,8){\line(2,-1){10}}
\put(80,3){(b)}
\end{picture}
\vskip 20pt

\begin{picture}(140,30)
\put(5,15){\line(1,0){15}}
\put(5,16){\line(1,0){15}}
\put(20,15){\line(1,1){10}}
\put(20,16){\line(1,-1){10}}
\put(20,15){\line(1,0){15}}
\put(20,16){\line(1,0){15}}
\put(35,15){\line(1,1){10}}
\put(35,16){\line(1,-1){10}}
\put(27,12){$\rho$}
\put(15,3){(c)}

\put(70,15){\line(1,0){15}}
\put(70,16){\line(1,0){15}}
\put(85,15){\line(1,1){10}}
\put(85,16){\line(1,-1){18}}
\put(94,7){\line(1,0){15}}
\put(93,8){\line(1,0){16}}
\put(109,7){\line(1,1){10}}
\put(109,8){\line(1,-1){10}}
\put(101,4){$\rho$}
\put(85,3){(d)}
\end{picture}
\vskip 20pt

\begin{picture}(140,30)
\put(5,15){\line(1,0){15}}
\put(5,16){\line(1,0){15}}
\put(20,16){\line(1,-1){8}}
\put(20,15){\line(1,-1){8}}
\put(20,16){\line(1,1){8}}
\put(20,15){\line(1,1){8}}
\put(28,7){\line(2,1){10}}
\put(28,8){\line(2,-1){10}}
\put(28,23){\line(2,1){10}}
\put(28,24){\line(2,-1){10}}
\put(22,21){$\rho$}
\put(22,6){$\rho$}
\put(15,3){(e)}

\put(75,15){\line(1,0){15}}
\put(75,16){\line(1,0){15}}
\put(90,15){\line(1,1){10}}
\put(90,16){\line(1,-1){8}}
\put(90,15){\line(1,-1){8}}
\put(98,8){\line(1,-1){10}}
\put(98,8){\line(1,0){11}}
\put(98,7){\line(1,0){11}}
\put(109,8){\line(1,-1){10}}
\put(109,7){\line(1,1){10}}
\put(85,3){(f)}
\put(95,12){$a_{1}$}
\put(104,4){$\rho$}
\end{picture}
\vskip 20pt

\begin{picture}(140,30)
\put(5,15){\line(1,0){15}}
\put(5,16){\line(1,0){15}}
\put(20,15){\line(1,1){10}}
\put(20,16){\line(1,-1){8}}
\put(20,15){\line(1,-1){8}}
\put(28,7.5){\line(1,0){10}}
\put(28,7){\line(2,1){10}}
\put(28,8){\line(2,-1){10}}
\put(15,3){(g)}
\put(25,12){$a_{1}$}
\end{picture}

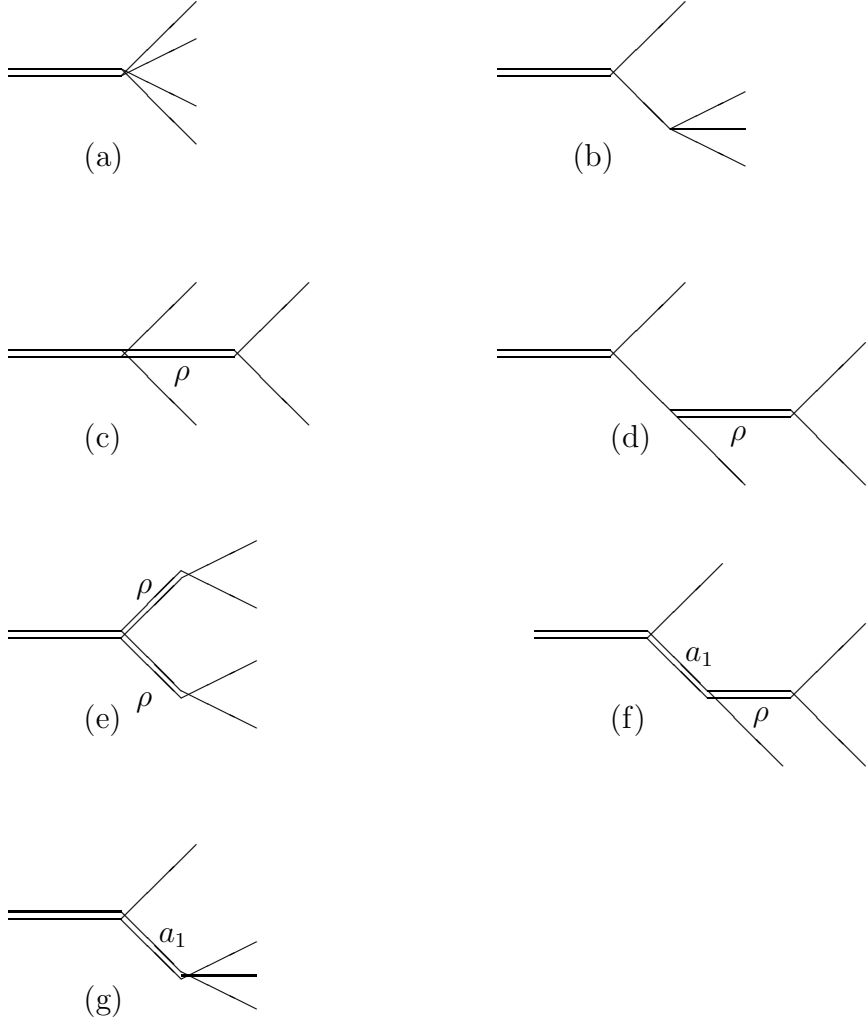
\noindent Figure 1. Diagrams contributing to the $\rho^0\to 4\pi$ decays in
chiral effective Lagrangians of $\pi$, $\rho$ and $a_1$ mesons. Single lines
denote pions, double lines spin-1 mesons. Anomalous terms in the effective
Lagrangian introduce diagrams similar to (f) and (g) but with an $\omega$
meson replacing the $a_1$.
\end{document}